\documentclass[prmaterials,onecolumn,superscriptaddress,floatfix,showpacs,amsmath,amssymb]{revtex4}

\usepackage{graphicx}
\usepackage{dcolumn}
\usepackage{hhline}
\usepackage{bm,color}
\usepackage{float}
\usepackage{amsmath}
\usepackage{epstopdf}

\usepackage{amsmath}
\usepackage{times}
\usepackage{color}
\usepackage{amsmath}
\usepackage{amssymb}
\usepackage{tabularx}

\newcommand{\cto}{CoTiO$_{3}$}
\newcommand{\nto}{NiTiO$_3$}

\newcommand{\tn}{$T_{\rm N}$}

\newcommand{\mb}{$\mu_{\rm B}$}

\newcommand{\etal}{\textit{et~al.}}

\begin{document}

\title{Magnetostriction and magnetostructural domains in CoTiO$_3$}

\author{K.~Dey}
\email[email:]{kaustav.dey@kip.uni-heidelberg.de}
\affiliation{Kirchhoff Institute of Physics, Heidelberg University, INF 227, D-69120, Heidelberg, Germany}

\author{M.~Hoffmann}
\affiliation{Kirchhoff Institute of Physics, Heidelberg University, INF 227, D-69120, Heidelberg, Germany}

\author{R.~Klingeler}
\affiliation{Kirchhoff Institute of Physics, Heidelberg University, INF 227, D-69120, Heidelberg, Germany}
\affiliation{Center for Advance Materials (CAM), Heidelberg University, INF 225, D-69120, Heidelberg, Germany}

\date{\today}

\begin{abstract}

We report the magnetostrictive length changes of \cto\ studied by means of high-resolution dilatometry in magnetic fields ($B$) up to 15~T. In the long-range antiferromagnetically ordered phase below \tn\ = 38~K, the easy-plane type spin structure undergoes a spin-reorientation transition in the $ab$ plane in magnetic fields $B || ab \approx 2$~T. We observe pronounced length changes driven by external magnetic field in this field region indicating significant magnetoelastic coupling in \cto. Specifically, we observe anisotropic deformation of the lattice for fields applied in the $ab$ plane. While, for $B \lesssim 2$~T, in-plane magnetostriction shows that the lattice expands (contracts) parallel (perpendicular) to the field direction, the opposite behaviour appear at higher fields. Furthermore, there are remarkable effects of slight changes in the applied uniaxial pressure on the magnetostrictive response of \cto\ persisting to temperatures well above \tn. The data evidence the presence of magnetic domains below \tn\ as well as of structural ones in \cto. The presence of magnetic domains in the spin ordered phase is further evidenced  by an additional 3-fold magnetic anisotropy appearing below \tn. We discuss the effects of rotational magnetic domains on isothermal magnetization and magnetostriction and interpret our results on the basis of a multi-domain phenomenological model.

\end{abstract}
		
\maketitle

 \section{Introduction}

Research in antiferromagnetic spintronics has garnered enormous interest over the past decades due to its high-frequency (THz) dynamics, long distance spin-wave transport, spin-orbit effects and robustness against external magnetic field perturbations, making them attractive for practical applications (see Refs. [\onlinecite{Gomonay_2020,JUNGFLEISCH2018}] and references therein). In particular, considerable efforts have been made to understand and manipulate the magnetic domain structure \cite{teo_2020}, spin-wave scattering at the domain walls \cite{Ross2020} and domain wall motion \cite{stuart-2015,shen_2020}. Understanding the domain wall structure and dynamics is not only relevant from an application point of view, but also has profound effects on the macroscopic thermodynamics and transport properties of strongly correlated electron systems~\cite{Gomonaj_1999, Kalita_2005, Nafradi_2016, Dagotto2005, dey_2021}.

\begin{figure}[h]
\centering
\includegraphics[width=1\columnwidth]{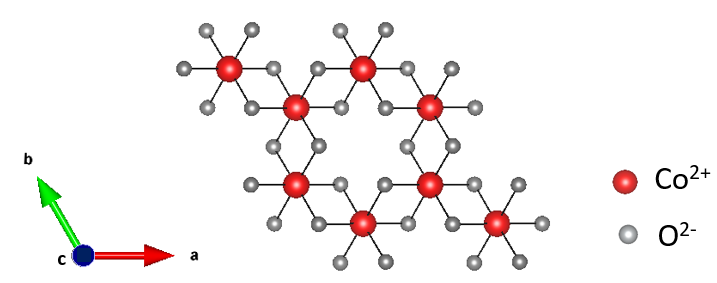}
\caption {Crystallographic arrangement of cobalt and oxygen ions in a basal hexagonal $ab$ plane of the ilmenite \cto\ structure \cite{icsd2}. The buckling is along the $c$ axis and hence is not visible here.}
\label{struct}
\end{figure}

In this report, we investigate \cto\ belonging to the ilmenites titanates family ($M$TiO$_3$; $M =$ Co, Ni, Mn, Fe), which have been studied quite extensively in the recent years due their significant magnetoelectric and magnetoelastic properties~\cite{Harada2016,Dubrovin_2020,Dey2020,dey_2021, mufti-2011,charilou_2012}. All the ilmenite titanates crystallize in the rhombohedral $R\bar{3}$ structure with the magnetic $M^{2+}$ ions in the basal $ab$ plane arranged in a buckled honeycomb-like structure (see Fig.~\ref{struct}). In particular for \cto, early magnetization \cite{Watanabe_1980, Balbashov2017} and neutron diffraction experiments \cite{Newnham_1964} indicated a long-ranged easy-plane type antiferromagnetic structure below \tn\ = 38~K. Furthermore, recent inelastic neutron scattering experiments on \cto\ exhibited the presence of Dirac magnons~\cite{Yuan_2019} and rendered it as a model system to study non-trivial magnon band topology~\cite{elliot2020,Yuan2020}. Here, we present macroscopic magnetization and magnetostriction ($dL(B)/L$) measurements on \cto. In particular, we observe pronounced effects of rotational magnetic domains on the macroscopic measurements and interpret our results on the basis of multi-domain phenomenological theories. Also above \tn , the magnetostrictive response is extremely sensitive to small changes in the applied external uniaxial pressure, indicating the presence of magnetostructural domains in \cto.

\section{Experimental Methods}

Macroscopic single crystals of \cto\ were grown by the optical floating-zone technique as reported elsewhere~\cite{cryst}. Static magnetic susceptibility $\chi=M/B$ was studied in magnetic fields up to 15~T applied along the principal crystallographic axes by means of a home-built vibrating sample magnetometer~\cite{klingeler_2006} and in fields up to 5~T in a Quantum Design MPMS-XL5 SQUID magnetometer. The angular dependence of magnetisation was measured in a Quantum Design MPMS3 SQUID magnetometer using the rotator option. The relative length changes $dL_i/L_i$ were studied on a single crystal of approximate dimensions $1.6 \times 2.0 \times 1.3~$mm$^{3}$ by means of a standard three-terminal high-resolution capacitance dilatometer ~\cite{Kuechler2012,LiranWang2018}. The field induced uniaxial length changes $dL_i(B)/L_i$ were measured at various fixed temperatures in magnetic fields up to 15~T and the longitudinal magnetostriction coefficients $\lambda_i = 1/L_i\times dL_i(B)/dB$ were derived with the same dilatometer. In both cases, magnetic fields were applied along the direction of measured length changes. To further study the length changes transverse to applied external magnetic fields, a second setup comprising of a mini-dilatometer on a rotator~\cite{Kuchler2017} inside a Physical Properties Measurement System (Quantum Design) was applied.

\section{Results}

\begin{figure}[h]
\centering
\includegraphics[width=0.9\columnwidth]{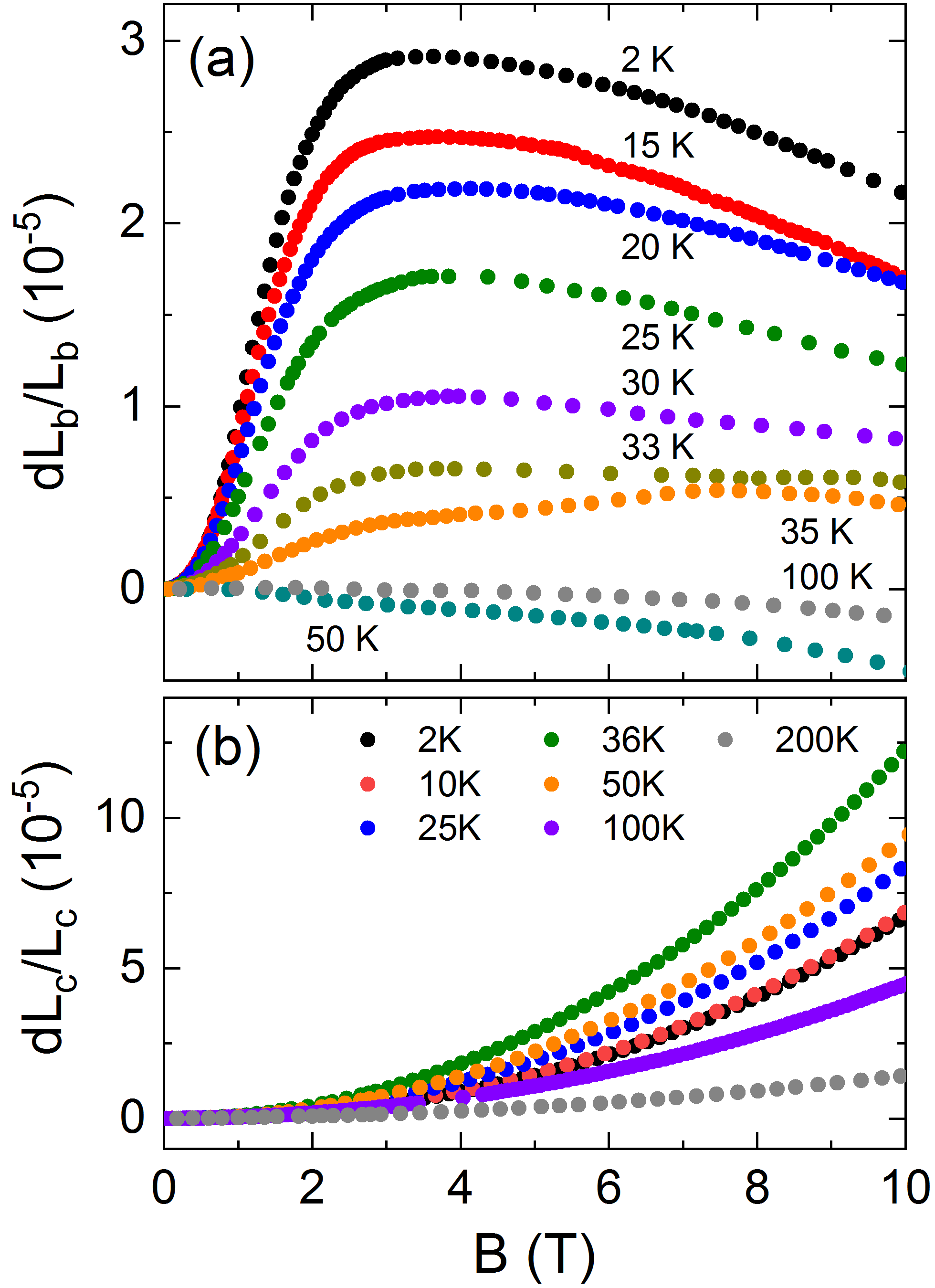}
\caption {Relative length changes $dL_i/L_i$ versus magnetic field for (a) $B || b$ (2.5~MPa) and (b) $B || c$ at different temperatures.}
\label{fig:MS_b}
\end{figure}

\begin{figure}[htb]
\centering
\includegraphics[width=\columnwidth]{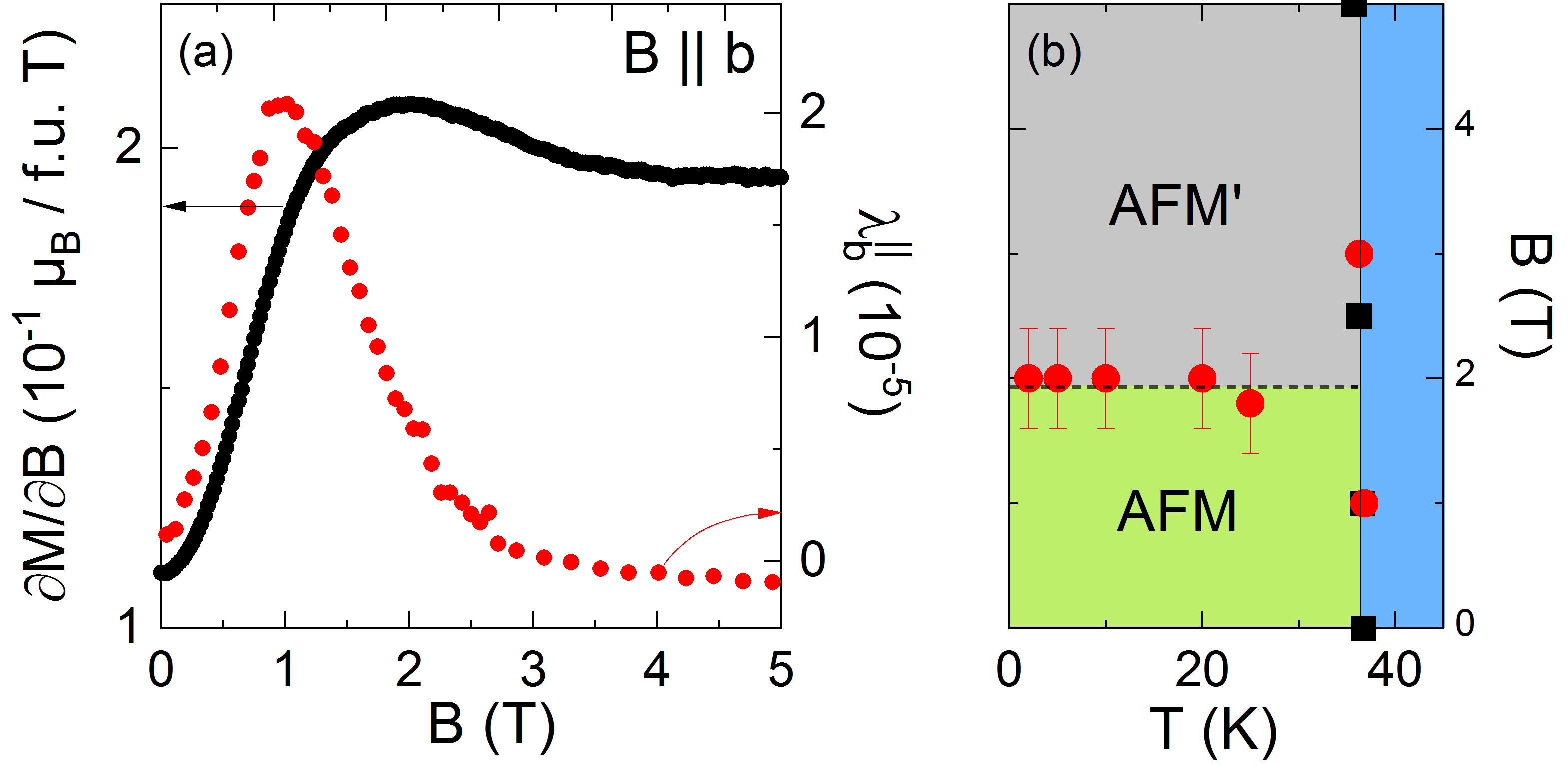}
\caption {(a) Field derivative of the isothermal magnetization $\partial M/\partial B$ and magnetostriction coefficient $\lambda_b$ for $B||b$ axis, at 2~K. The maxima of both quantities are found at different fields. (b) Magnetic phase diagram for $T\leq 42$~K and $B$ up to 5~T. Red (black) markers represent positions of magnetisation (thermal expansion) anomalies. Lines and colours are guide to the eye.}
\label{fig:lambda_dmdb}
\end{figure}

The application of external magnetic-field yields pronounced changes in the lattice parameters of \cto. As shown in Fig.~\ref{fig:MS_b}(b), a monotonic increase of the $c$ axis  driven by $B || c$ up to 15~T is observed at all the measured temperatures, i.e., between 2 and 200~K. In contrast, for $B || b$ a pronounced increase of the associated lattice parameter is observed in the low-field region, i.e., $B < 3$~T, followed by a relatively smaller decrease in higher fields as seen in Fig.~\ref{fig:MS_b}(a). For $T\lesssim T_{\rm N}$, small anomalies in the magnetostriction (i.e., in the data obtained at 33 and 35~K) at higher field indicate the phase boundary to the paramagnetic phase (cf. also Fig.~\ref{fig:lambda_dmdb}(b) and Fig. \ref{fig:MS_hyst}).

The region of significant lattice changes below 3~T applied in-plane coincides with the non-linear region in magnetization $M$ vs. $B$ measurements~\cite{Balbashov2017}, signalling a spin-reorientation process in the $ab$-plane. This is particularly evident in Fig.~\ref{fig:lambda_dmdb}(a) where the corresponding derivatives of both, i.e., of the magnetization $\partial M/\partial B$ and of the relative length changes $\lambda$ are presented. However, these quantities peak at different positions which is contrary to the expectations of bare spin-orientation associated with an anomaly in the magnetostriction due to finite spin-lattice coupling~\cite{Rotter_M, Ackermann_2015, Werner_2017}. Instead, it is reminiscent of rotational domain behaviour as observed in Ising-like BaCo$_2$V$_2$O$_8$~\cite{Niesen_2013} or in easy-plane type BaNi$_2$V$_2$O$_8$~\cite{Knafo_2007} and \nto ~\cite{dey_2021}. As described below, the presence of domains has to be involved and in the following we will present clear evidence that the data represent a crossover from a low-field multi-domain state to a high-field homogeneous state with spins nearly perpendicular to the applied magnetic field.

\begin{figure}[tbh]
\centering
\includegraphics[width=1 \columnwidth]{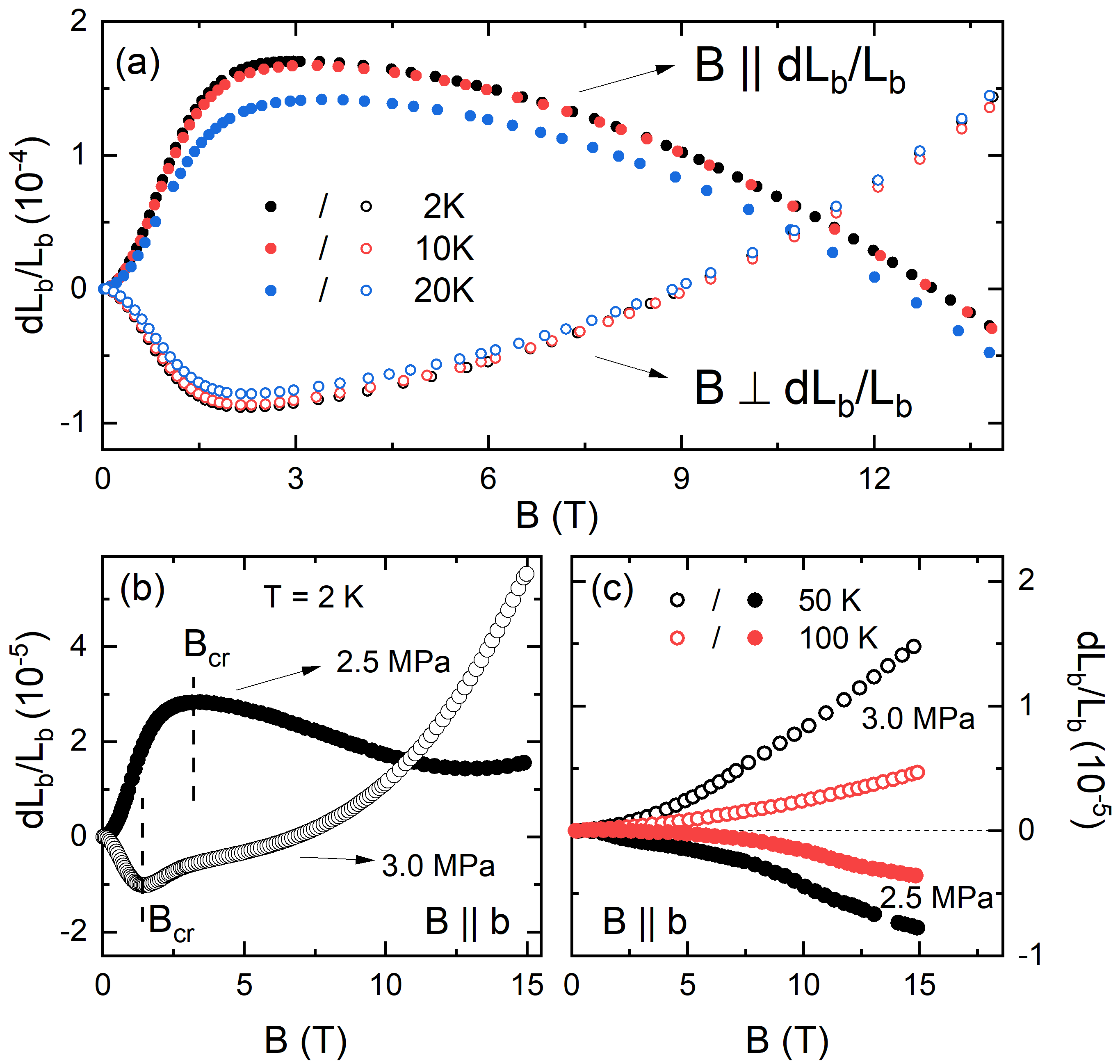}
\caption {(a) Magnetostriction $dL_b/L_b$ versus magnetic field $B || b$ and $B \perp b$ (both in-plane), respectively, at different temperatures. (b) $dL_b/L_b$ vs. $B$, at $T = 2$~K, with different uniaxial pressure applied along the $b$ direction, and (c) in the paramagnetic phase at 50~K and 100~K, respectively. The vertical dashed lines in (b) indicates approximately the crossover from magnetic multi-domains to the monodomain region. See the text for details.}
\label{magnetostruct_dom}
\end{figure}

The lattice changes in the hexagonal $ab$ plane are further investigated by measuring the length changes $dL_b/L_b$ in longitudinal and transversal magnetic fields applied in-plane. As shown in Fig.~\ref{magnetostruct_dom}(a), the longitudinal length changes ($dL_b(B||b)/L_b$) exhibit an almost equal and opposite behaviour as compared to transverse length changes where the field is applied $\perp bc$, i.e., along the perpendicular-to-$b$ in-plane direction ($dL_b(B\perp bc)/L_b$). Specifically, while above 3~T the spins are nearly perpendicular to $B$, the data imply an anisotropic deformation of the $ab$ plane. This behaviour does not comply to the perfect $R\bar{3}$ structure but reveals an in-plane anisotropy induced by applied external magnetic field. The field-direction induced deformation can be estimated as the difference in magnetostrictive length changes for $B||b$ and $B\perp b$ which, at 3~T, amounts to $\approx 2.6 \times 10^{-4}$ at $T =2$~K. Tentatively, as the field lifts the hexagonal lattice symmetry and rotates spins perpendicularly to $B$, this quantity may be interpreted as an orthorhombic distortion. Above the spin-crossover region, i.e., at $B \geq 3$~T, the data reveal that the hexagonal basal plane shrinks parallel to the field-direction (i.e., perpendicular to the spin direction), whereas it elongates perpendicular to the field direction (parallel to the spin direction).

In a capacitance dilatometer, uniaxial pressure amounting to a few MPa is applied to the sample during magnetostriction measurements resulting in very small anisotropic strain~\cite{Scaravaggi2021},~\footnote{Magnetisation studies as shown in Fig.~\ref{fig:lambda_dmdb} naturally do not employ such uniaxial strain.}. Since pressures in the GPa range are required to induce significant elastic deformation in solids, such tiny distortions inevitably associated with the measurement setup are usually irrelevant. Notably, for \cto\ we observe quite remarkable effects of tiny pressures applied by the dilatometer in magnetostriction measurements. As seen in Fig.~\ref{magnetostruct_dom}(b) and (c), an increase in the uniaxial pressure by mere 0.5~MPa applied by tightening the dilatometer screw results in two significant changes in the magnetostrictive behaviour: (1) there is a complete reversal in the $dL_b/L_b$ vs.~$B$ behaviour in the whole field range up to 15~T under study and up to 100~K and (2) a shift in the spin crossover field $B_{cr}$ marked by the vertical dashed lines in Fig.~\ref{magnetostruct_dom}(b) to lower magnetic fields. Note, that the measurements are performed in a single-run without removing the sample. We argue below that observations (1) and (2) indicate the presence of magnetostructural domains in \cto\ single crystal~\cite{Knafo_2007,Niesen_2013}.

In particular at $T = 2$~K and $B > 5$~T, a closer look at the data in Fig.~\ref{magnetostruct_dom}(b) reveals a switch from lattice contraction to expansion upon increasing the uniaxial pressure from 2.5~MPa to 3.0~MPa. This pressure dependent reversal effect persists up to 100~K (see Fig.~\ref{magnetostruct_dom}(c)) which is well above \tn, suggesting a purely crystallographic origin. Interestingly, such pressure dependent reversal effect in the $dL_b/L_b$ behaviour has been previously observed in other systems and ascribed to the presence of crystallographic structural twins~\cite{Niesen_2013}. It is to note that, the recent neutron diffraction studies on \cto~\cite{elliot2020} confirm the presence of two almost equal-weighted structural twins related by a [110] mirror plane. Rigorous X-ray Laue pattern investigations on our \cto\ single crystals excludes the possibility of more than one crystallographic grain. Hence, the findings above leads to the conclusion that, the observed change from positive to negative magnetostriction upon increasing the uniaxial pressure in \cto, is due to the presence of structural twins. Consequently, the experimentally measured length changes reflect the superposition response from all the individual twins whose relative population is altered by the applied external uniaxial pressure, resulting in a complete reversal in $dL_b/L_b$ behaviour at a particular pressure.

\begin{figure}[tbh]
\centering
\includegraphics[width= \columnwidth]{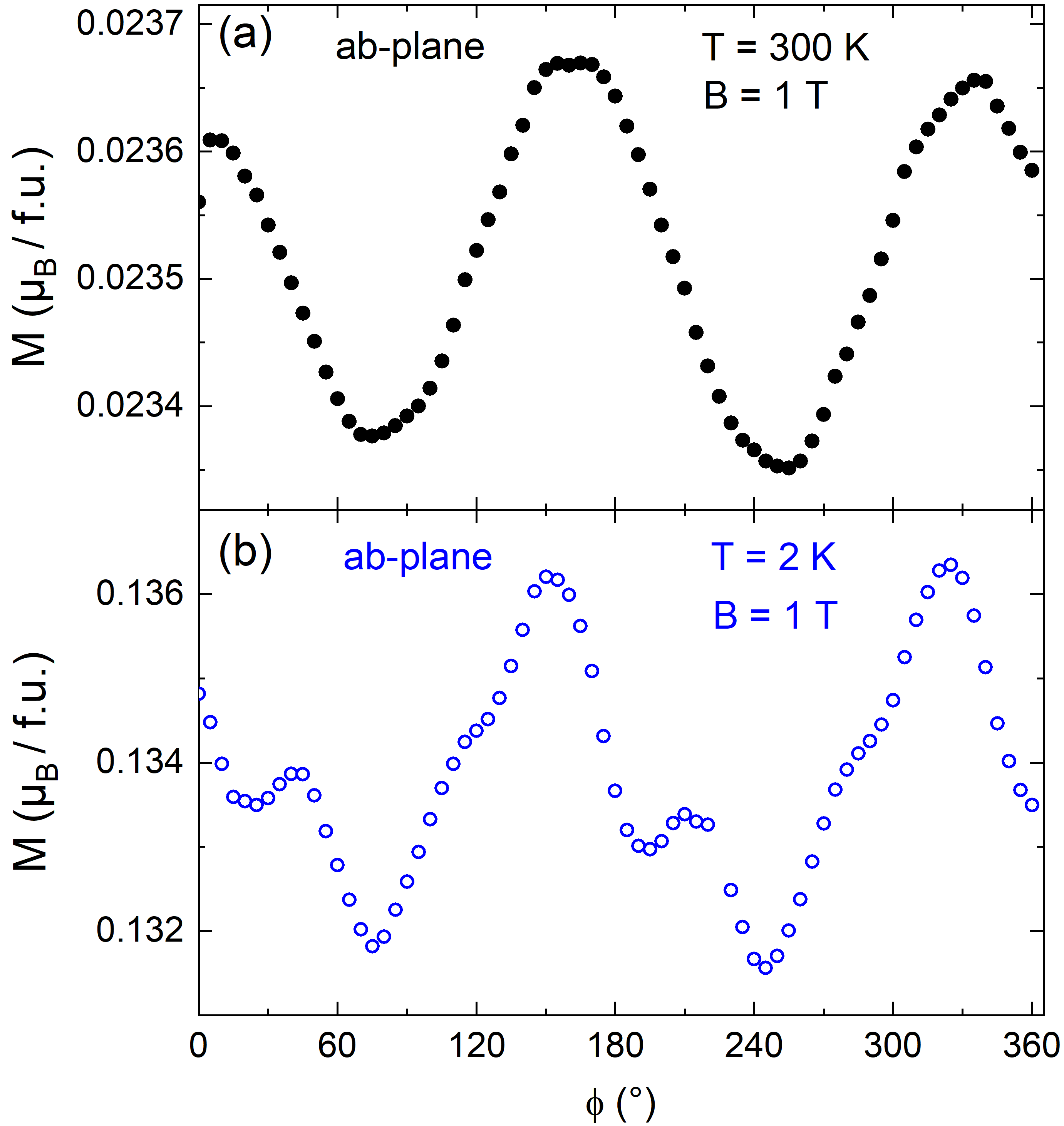}
\caption {The angle dependence of the magnetization at $B = 1$~T obtained by rotating the single crystal in the $ab$ plane at (a) 300~K and at (b) 2~K. An additional asymmetry observed in the magnetic ordered phase is interpreted as a signature of magnetic domains.}
\label{fig:M-vs-phi}
\end{figure}

In addition to structure domains evidenced above, the low-temperature spin ordered phase in principle may host magnetic domains, too. Specifically, the crystallographic symmetry is $C_3$ in the easy-hexagonal plane, which may result in the presence of three equally populated magnetic domains with spins rotated by $120 ^{\circ}$ between them. Such a domain structure is often found in easy-plane type antiferromagnets~\cite{Wilkinson_1959, Knafo_2007} and more recently in \nto, which is isostructural to \cto~\cite{dey_2021}. The presence of magnetic domains below \tn\ is indeed evidenced by the angle-dependent magnetization measurements $M$ vs. $\theta$ in the $ab$ plane, as exhibited in Fig.~\ref{fig:M-vs-phi}. In the paramagnetic region, i.e., at 300~K, the data reveal a $C_2$ (180$^\circ$) anisotropy which amounts to $\approx 2 \times 10^{-4}$~\mb/f.u. in the plane. In contrast, as seen in Fig.~\ref{fig:M-vs-phi}(b), we observe additional peaks in the angle dependence of the magnetisation in the magnetically ordered phase at $T$ = 2~K $<$ \tn. Tracking the maximum position of the peaks reveals an approximate 60$^{\circ}$ spacing between each of them which is indicative of a $C_3$ anisotropy in the easy $ab$-plane. We conclude the presence of a magnetic multi-domain ground state in \cto. Fingerprints of magnetic domains are also visible in the magnetostriction data as will be further discussed below.

\section{Discussion}

As shown in Fig.~\ref{fig:lambda_dmdb}, reorientation of the spins around $B\approx 2$~T is associated which lattice changes which derivative peaks well below the actual spin crossover field. Moreover, the characteristic field scale of the magnetostrictive lattice changes and even the sign of magnetostriction drastically depends on tiny changes of the strain applied in the experiment. These findings imply the presence of structural twins well above \tn\ and further support the scenario of magnetic domains at low temperatures. It also implies that the anomalies in the magnetostriction for $B||b$ cannot be analysed in terms of thermodynamic relations.

In contrast, magnetostriction along the $c$ axis is not affected by the presence of in-plane twins. $dL_{\rm c}(B)$ exhibits a monotonic field-driven increase which is an order of magnitude larger than the in-plane length changes (cf. Fig.~\ref{fig:MS_b}). Notably, for a linear in-field magnetization, i.e., $M = \chi B$, which is realised for the $c$ axis~\cite{Watanabe_1980,Dey2020} and in general for the paramagnetic region, exploiting the Maxwell relation $ 1/L_i\times (\partial L_i/ \partial B)\rvert_p = - 1/V \times (\partial M/ \partial p_i)\rvert_B$ results in $(dL_i/L_i)\rvert_p = -(1/2V) (\partial \chi/ \partial p_i)\rvert_B \times B^2$, where $V$ denotes the molar volume. Indeed, for $B||c$ the data in Fig.~\ref{fig:MS_b}(b) and SI Fig.~1 confirm the square-field dependency $dL_c/L_c \propto B^2$ for $B \leq 15$~T. Furthermore, the data enable reading off $\partial (\ln \chi_c)/ \partial p_c \simeq -21$~\%~GPa$^{-1}$ at 2~K. For $dL_{\rm b}(B)$, in contrast, at 100~K, which is well in the paramagnetic regime, the Maxwell relation described above would imply a change in sign of $\partial \chi_b/\partial p_b$ upon increasing the pressure from 2.5 to 3.0~MPa which is unreasonable. Instead, this behaviour is associated to the presence of structural domains in-plane and the obtained magnetostrictive behaviour is a superposition of the response of two structural twins whose relative proportions are changed due to change in uniaxial pressure, resulting in the sign changes in $dL_b/L_b$ behaviour.

In the long-range spin ordered phase, increasing the uniaxial pressure in the few-MPa regime described above, shifts the crossover field $B_{cr}$ as observed in Fig.~\ref{magnetostruct_dom}(b). This is attributed to changes in the magnetic domain distribution of the ground state. Hence, as stated above and in-line with our expectations, the broad-hump observed in $dL_b/L_b$ (see Figs.~\ref{fig:MS_b}(a) and \ref{magnetostruct_dom}(b)) is not the signature of a thermodynamic phase transition but signals the crossover from the multi-domain spin ground state to a uniform phase comprising of spins pointing nearly perpendicular to the external magnetic field direction.

 \begin{figure}[t]
\centering
\includegraphics[width= \columnwidth]{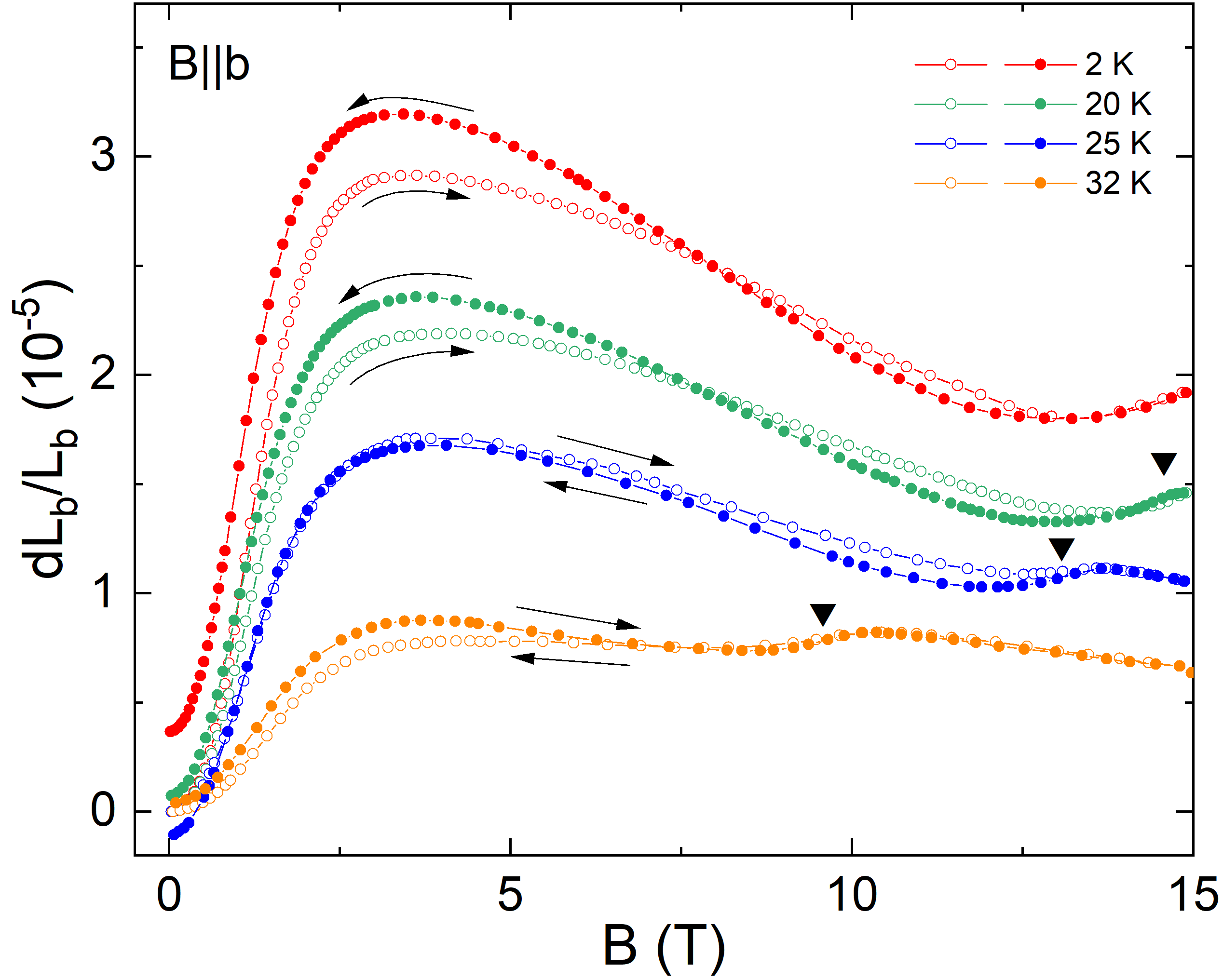}
 \caption {Magnetostriction along the $b$ axis during imposition (open markers, up-sweep) and removal (full markers, down-sweep) of the magnetic field at different temperatures and at an estimated uniaxial pressure of 2.5~MPa. Small discontinuities in the magnetostriction at higher fields indicated by the black triangles signal the antiferromagnetic to paramagnetic phase boundary.}
\label{fig:MS_hyst}
\end{figure}

Interestingly, the up- and down-sweeps in magnetostriction feature a small remanent striction in the antiferromagnetic phase (see Fig.~\ref{fig:MS_hyst}). This observation further evidences the presence of antiferromagnetic domains as it is straightforwardly explained by irreversibility due to the pinning of domain walls by defects and crystalline imperfections~\cite{Gomonaj_1999}. The remanent striction amounts to $3.7 \times 10^{-6}$ at 2~K and 2.5~MPa. The small hysteresis at higher fields has an opposite sign as compared to the low-field one and might indicate either the presence of few strongly pinned domain walls persisting in the whole antiferromagnetic phase or a weak discontinuous character of the antiferromagnetic phase boundary at high fields.

\begin{figure}[h]
\centering
\includegraphics[width=\columnwidth]{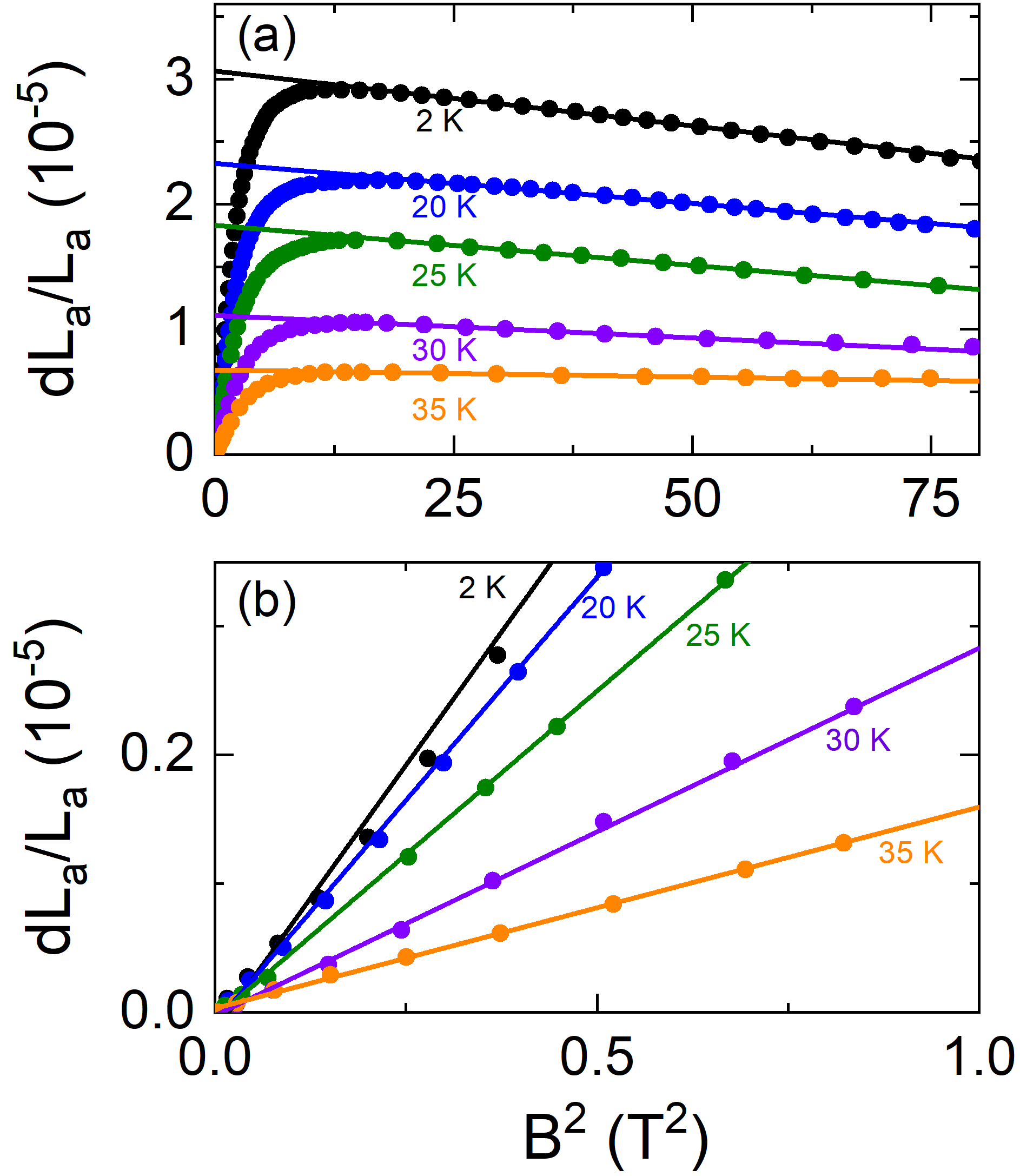}
\caption {$dL_b/L_b$ versus $B^2$ in the (a) high-field homogeneous state and (b) low-field multi-domain state, respectively. The solid lines are corresponding linear fits using Eqs.~(1) and (2), respectively. See the text for details.}
\label{domaintheory}
\end{figure}

The overall scenario is, therefore, that application of external magnetic fields leads to rotation of spins facilitated by domain wall motion. In the multi-domain ground state, the macroscopic magnetization and relative length changes in magnetic fields ($B \lesssim 3$~T) undergo continuous changes and are characterized by broad anomalies in the corresponding field derivatives which maxima however do not coincide (see Fig.~\ref{fig:lambda_dmdb}). Such behaviours have been previously observed in several easy-plane type, high-symmetry cubic~\cite{Safa_1978,Tanner_1979} and hexagonal antiferromagnets~\cite{Knafo_2007, Kalita_2000,Kalita_2005,dey_2021}. We hence in the following apply the phenomenological models developed by Kalita \etal~\cite{Kalita_2000,Kalita_2001,Kalita_2002, Kalita_2004, Kalita_2005} to describe magnetization and magnetostriction in \cto. Both in the multi-domain region $B < 1$~T and in the homogeneous state at $B > 3$~T, the induced striction $dL_b/L_b$ varies as a square of magnetic field as shown in Fig.~\ref{domaintheory}. At higher fields, i.e., in the homogeneous phase characterized by spins perpendicular to the applied field direction, the magnetostriction is described well by~\cite{Kalita_2005}

\begin{equation}
\left (\frac{dL_b}{L_b} \right) (T, B) = \left (\frac{dL_b}{L_b} \right)_0(T) + \alpha(T) \cdot B^2
\end{equation}

where $(dL_b/L_b)_0$ is the hypothetical spontaneous magnetostriction that would be observed if the magnetoelastic domains did not appear at low fields; $\alpha(T)$ is a temperature-dependent prefactor.

The large magnetostrictive response in the multi-domain region is facilitated by domain effects and can be described by \cite{Kalita_2005}

\begin{equation}
 \left (\frac{dL_b}{L_b} \right) (T, B) =  \left (\frac{dL_b}{L_b} \right)_0(T) \times \left (\frac{B}{B_d} \right)^2
\end{equation}

where $B_d$ is a phenomenological parameter. Note, that the slope of the $(B/B_d)^2$-behaviour is again the hypothetical spontaneous magnetostriction $(dL_b/L_b)_0$ which is accessible from fitting the experimental data using Eq.~(1). As seen clearly in Fig.~\ref{domaintheory}, the equations (1) and (2) describe the $ab$ plane $dL(B)/L$ data well in both the multi-domain and the homogeneous state. We note that, while Fig.~\ref{domaintheory} shows the analysis of magnetostriction data obtained at the external pressure of 2.5~MPa, the $B^2$-behaviour is also found at different external pressure (cf. Fig.~\ref{magnetostruct_dom}(b)), in both regimes $B < 0.5$~T and $B > 3$~T, respectively.


\begin{figure}[h]
\centering
\includegraphics [width=1\columnwidth] {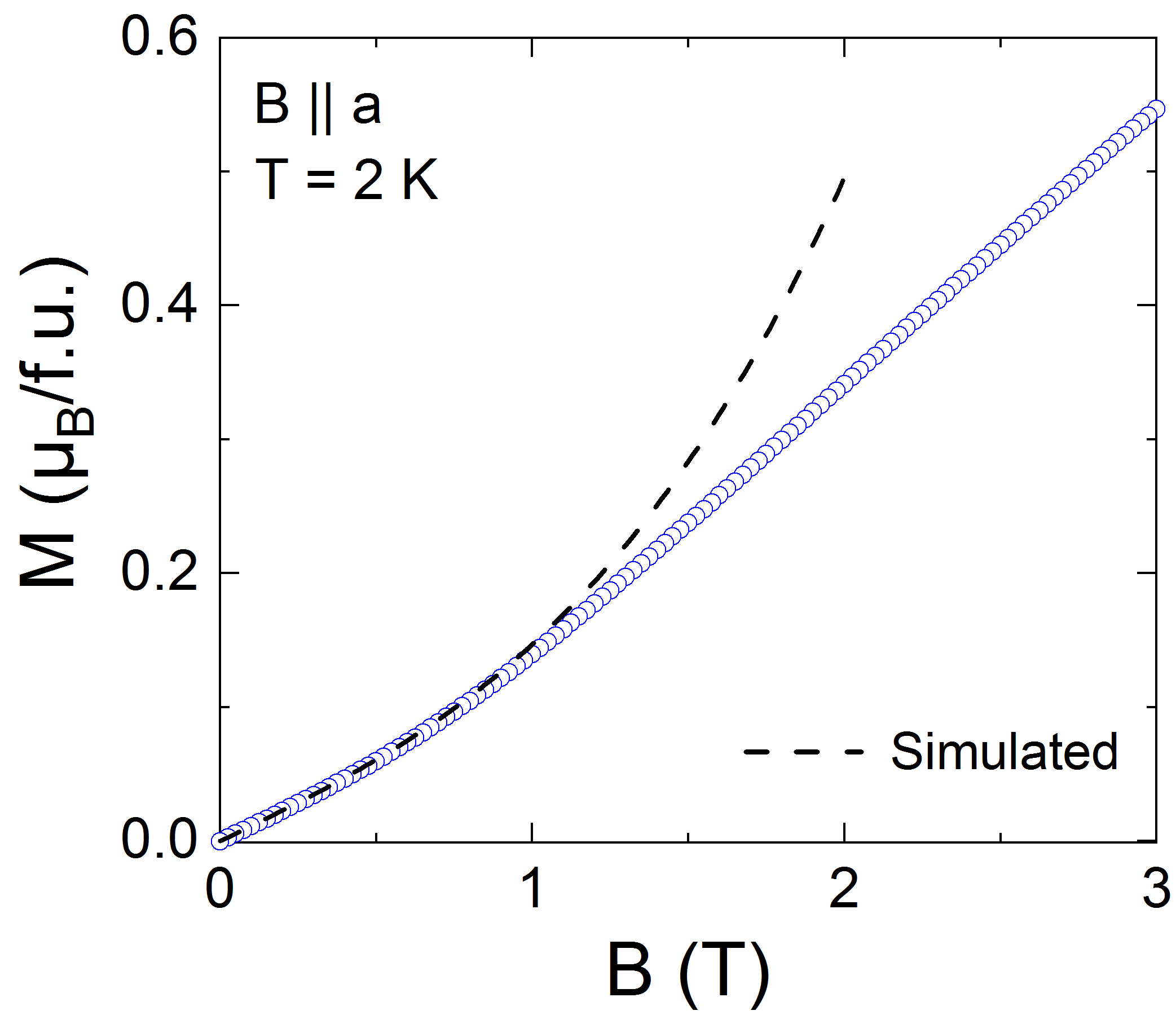}
\caption{Magnetization $M$, at $T=2$~K, versus applied magnetic field $B||a$-axis. The dashed black line shows a simulation to $M$ at low fields (see the text for more details).}\label{simM}
\end{figure}


Following the phenomenological models, a smooth non-linear (sickle-shaped) variation of $M(B)$ is expected at low magnetic fields. It is a characteristic of a multi-domain ground state where spin-reorientation gradually takes place by displacement of domain walls~\cite{Tanner_1979}. It may be described by~\cite{Kalita_2005}
\begin{equation}
\label{M_B_calculated}
    M(B) = \frac{1}{2} \chi_e B \left( 1 + \frac{(dL_b/L_b)_{rem}}{(dL_b/L_b)_0} + \frac{B^2}{B_d^2} \right).
\end{equation}

where $\chi_e$ is the magnetic susceptibility in the high-field linear (homogeneous phase) region. A linear fit to the $M$ vs $B$ data at $B > 4$~T yields $\chi_e = 0.19(1)$~\mb/ T f.u.. We note that Eq.~\eqref{M_B_calculated} does not include any free fitting parameters. Using $B_d$, $(dL_b/L_b)_{rem}$ and $(dL_b/L_b)_{0}$ from the analysis of the magnetostriction data, we simulate the field dependence of $M$ in the low-field region. As seen in Fig.~\ref{simM}, the simulated black line using Eq.~\eqref{M_B_calculated} yields a good description of magnetization up to 0.7~T, thereby further confirming the applied phenomenological model.

\section{Conclusions}

To summarize, we report the magnetostriction of \cto\ single crystals in magnetic fields up to 15~T obtained by high-resolution dilatometry experiments. The results evidence structural domains at high temperatures as well as magnetic domains in the long-range antiferromagnetically ordered phase. Below \tn , we observe pronounced magnetostrictive length changes corresponding to spin-reorientation in the $ab$ plane. Upon application of an in-plane magnetic field, there is an anisotropic deformation breaking the rhombohedral $R\bar{3}$ symmetry in the $ab$ plane. At $B=3$~T, the difference of in-plane lattice distances amounts to $2.6\times 10^{-4}$ which is hardly accessible by diffraction studies. Remarkably, there are drastic effects of small external uniaxial pressure on the magnetostrictive response both above and below \tn\ which is attributed to the presence of rotational magnetic and twin domains in \cto. We discuss the effects of domains on the magnetisation and magnetostriction on the basis of multi-domain phenomenological models. While the validity of such models and in particular the observed gigantic pressure effects on the magnetostriction are clear fingerprints of magnetostructural domains, our data also imply that even the sign of macroscopic magnetostriction in such materials hosting a multi-domain ground state can depend on tiny changes of the external parameters such as uniaxial strain.

\section*{Acknowledgements}
This work has been performed in the frame of the International Max-Planck School IMPRS-QD. We acknowledge financial support by the BMBF via the project SpinFun (13XP5088) and by the Deutsche Forschungsgemeinschaft (DFG) under Germany’s Excellence Strategy EXC2181/1-390900948 (the Heidelberg STRUCTURES Excellence Cluster) and through project KL 1824/13-1.


\end{document}